\documentclass[journal]{IEEEtran}
\IEEEoverridecommandlockouts %showing the thanks at the left bottom of the first page
\usepackage{epsfig,latexsym}
\usepackage{float}
\usepackage{indentfirst}
\usepackage{amsmath}
\usepackage{amssymb}
\usepackage{times}
\usepackage{graphicx}
\usepackage{ulem}
\usepackage{xcolor}
\usepackage{adjustbox}
\usepackage{footnote}
\usepackage{subfig}
\usepackage{titlesec}
\titlespacing\section{0pt}{12pt plus 4pt minus 2pt}{0pt plus 2pt minus 2pt}
\titlespacing\subsection{0pt}{12pt plus 4pt minus 2pt}{0pt plus 2pt minus 2pt}
\begin{document}

\title{BE-RAN: Blockchain-enabled Open RAN for 6G with DID and Privacy-Preserving Communication}
% \title{BE-RAN: Blockchain-enabled RAN with Decentralized Identity Management and Privacy-Preserving Communication for Open RAN and beyond}

\author{Hao~Xu,
        Zihan~Zhou,
        Lei~Zhang,~\IEEEmembership{Senior Member,~IEEE,}
        Yunqing~Sun,
        and~Chih-Lin~I,~\IEEEmembership{Fellow,~IEEE}% <-this % stops a space
\thanks{Hao Xu is with College of Electronics and Information Engineering, Tongji University, Shanghai, China, E-mail: hao.xu@ieee.org; Zihan Zhou is with Information Hub, Hong Kong University of Science and Technology (Guangzhou), E-mail: zzhou789@connect.hkust-gz.edu.cn;  Lei Zhang is with James Watt School of Engineering, University of Glasgow,
G12 8QQ, UK, e-mail: Lei.Zhang@glasgow.ac.uk; Yunqing Sun is with Department of Computer Science, McCormick School of Engineering and Applied Science, Northwestern University, Evanston, IL, US, E-mail: yunqing.sun@northwestern.edu; C. L. I is the Chief Scientist of Wireless Technologies at China Mobile Research Institute, Beijing, China, e-mail: icl@chinamobile.com}% <-this % stops a space
\thanks{Corresponding authors: Yunqing Sun}
\thanks{A study item is formulated by the authors in O-RAN Security Focus Group for potential standardization.}% <-this % stops a space

\thanks{}}

% \markboth{Journal of \LaTeX\ Class Files,~Vol.~14, No.~8, August~2015}%
% {Shell \MakeLowercase{\textit{et al.}}: Bare Demo of IEEEtran.cls for IEEE Journals}

\maketitle

\begin{abstract}
  As 6G networks evolve towards a synergistic system of Communication, Sensing, and Computing, Radio Access Networks become more distributed, necessitating robust end-to-end authentication. We propose Blockchain-enabled Radio Access Networks, a novel decentralized RAN architecture enhancing security, privacy, and efficiency in authentication processes. BE-RAN leverages distributed ledger technology to establish trust, offering user-centric identity management, enabling mutual authentication, and facilitating on-demand point-to-point inter-network elements and UE-UE communication with accountable logging and billing service add-on for public network users, all without relying on centralized authorities. We envision a thoroughly decentralized RAN model and propose a privacy-preserving P2P communication approach that complements existing security measures while supporting the CSC paradigm. Results demonstrate BE-RAN significantly reduces communication and computation overheads, enhances privacy through decentralized identity management, and facilitates CSC integration, advancing towards more efficient and secure 6G networks.

\end{abstract}

\begin{IEEEkeywords}
DLT, Blockchain, certificateless authentication, identity management, privacy preservation, Open RAN, O-RAN, SASE, SD-WAN, MPLS
\end{IEEEkeywords}

\section{Introduction}
\IEEEPARstart{A}{s} we move towards 6G networks, the traditional centralized architecture of cellular mobile networks, comprising Radio Access Networks (RAN) and Core Networks (CN), faces significant challenges in meeting the demands of emerging distributed applications such as industrial IoT, autonomous driving, and Industry 4.0 \cite{Chih-Lin2014, Yu2020}. These applications require not only lower latency and enhanced data privacy but also a synergistic system of Communication, Sensing, and Computing (CSC) that can satisfy both network operational needs and diverse user requirements \cite{Saad2020}. The current architecture, with its remote, centralized CNs, struggles to provide these capabilities effectively.
To address these issues, industries are adopting private 5G networks and Mobile Edge Computing (MEC) \cite{Chih-Lin2020, Rostami2019}. However, these solutions often involve high costs and potential privacy concerns for public users, particularly in terms of end-to-end authentication and secure identity management.
Concurrently, RAN is undergoing an architectural transition towards 6G New Radio and Open RAN initiatives, moving from physically distributed base stations to centrally managed, loosely coupled control functions (CU, DU, RU) \cite{3GPP2018, O-RANAlliance2020}. This shift, exemplified by C-RAN and O-RAN, enables more flexible network management through Network Function Virtualization (NFV) \cite{Chih-Lin2016, VanRossem2015}. This evolution presents an opportunity to implement robust end-to-end authentication mechanisms that can meet the complex demands of future networks.
The functional split of RAN at lower layers opens up possibilities for distributed features that could enhance privacy and security in decoupled RAN architectures. In this context, the adoption of distributed ledger technology in RAN presents an opportunity for implementing blockchain-native infrastructure \cite{Xu2020b, Yan2020, Tschorsch2016}. By integrating blockchain-enabled identity management and authentication functions within RAN, we can potentially provide more secure and user-centric services at a lower cost \cite{Chih-Lin2016}, while also facilitating the seamless integration of communication, sensing, and computing capabilities required for next-generation networks.

\subsection{Motivation}
\begin{figure}
  \includegraphics[width= 0.48\textwidth]{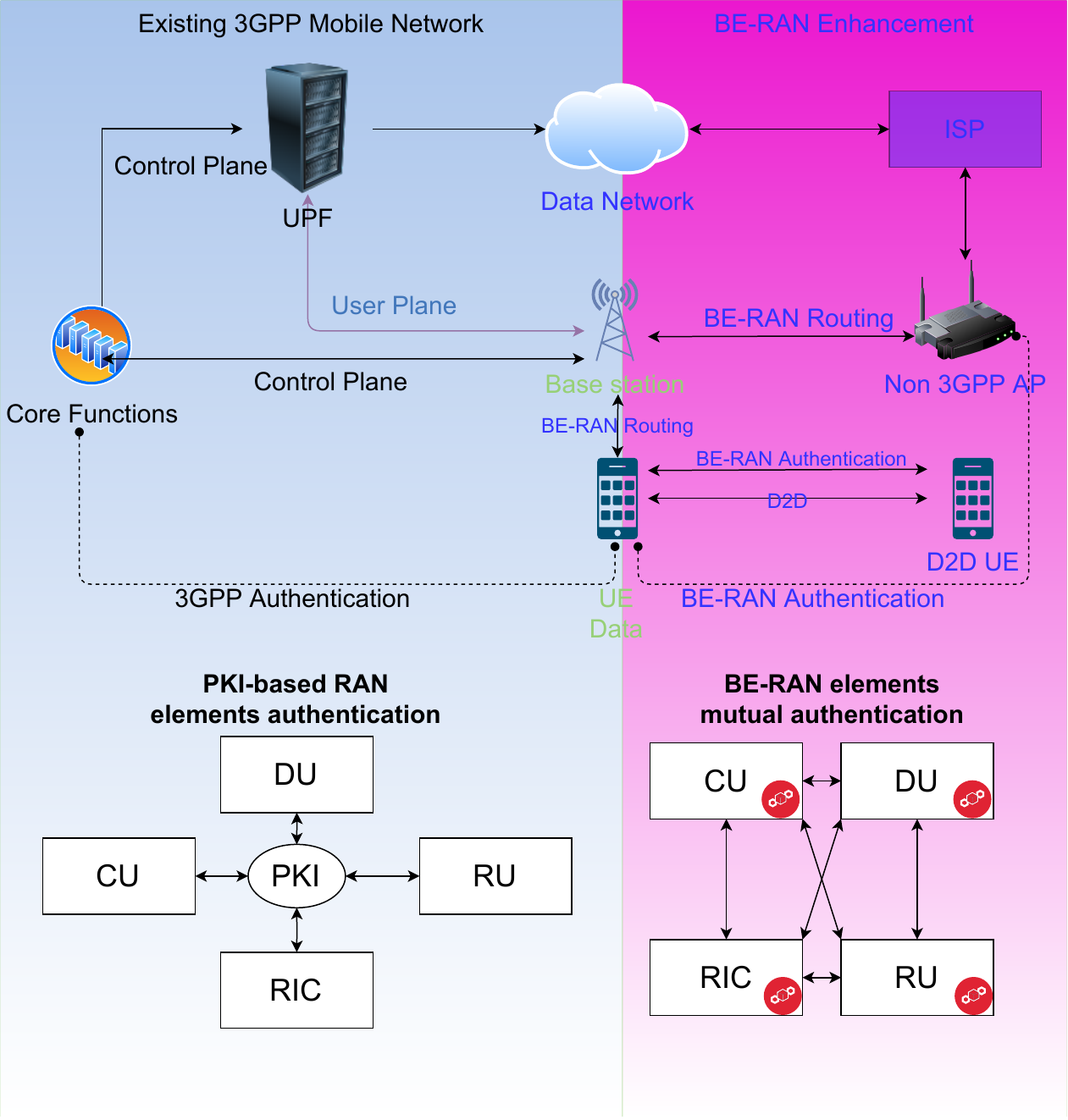}
  \caption{Overview of BE-RAN in addition to existing mobile network \label{fig:gateway}}

\end{figure}

The key benefit of decentralizing RAN functions is enabling privacy-preserving Point-to-Point (P2P) communication. This direct communication between users, without third-party interpretation, is crucial for ensuring privacy and security, especially in industrial settings that prioritize low latency and on-premise privacy.
Current cellular networks face challenges in independent P2P communication due to lack of global peer discovery and routing without compromising privacy and security. Additionally, the centralized architecture restricts direct communication between users in the same RAN cell, as user authentication, routing, and paging are only available at the Core Network (CN).
Distributed Ledger Technology (DLT) or Blockchain offers a solution by providing immutable, distributed records for user identities, enabling decentralized identity (DID) management and authentication. By implementing these functions locally at the RAN level with synchronized global identity records, users can communicate directly within the same RAN coverage without CN involvement. This approach improves privacy and reduces latency, as illustrated in Fig. \ref{fig:gateway}.
To address the growing demand for distributed communication in both industrial and consumer applications, we propose Blockchain-enabled Radio Access Network (BE-RAN). This system incorporates blockchain-enabled identity management and mutual authentication (BeMutual) as core functions, evolving RAN with a decentralization focus to provide enhanced privacy and connectivity for distributed scenarios.

\subsection{Contributions and Organization}

This paper presents a comprehensive BE-RAN architecture, introducing blockchain into RAN for revolutionary operational improvements. Our key contributions include:

\begin{itemize}
  \item A novel security framework for distributed RAN applications, enabling privacy-preserving D2D and peer-to-peer communications.
  \item A blockchain-enabled mutual authentication (BeMutual) architecture, combining zero-trust identity management without reliance on third-party CAs or PKI.
  \item Design guidelines for BE-RAN routing, switching, and QoS management, with a blockchain-based resource-sharing model.
\end{itemize}

BE-RAN supports decentralized communication scenarios (e.g., D2D, IIoT, autonomous driving), improving E2E latency and privacy. It offers user-centric identity management and on-demand point-to-point communication with accountable billing, enhancing resilience in emergencies.

The paper is organized as follows: Section \ref{sec:related} covers key concepts, Section \ref{sec:arch} introduces the BE-RAN framework, Section \ref{sec:func} details core security mechanisms, Section \ref{sec:results} provides performance analysis, Section \ref{sec:challenges} discusses challenges, and Section \ref{sec:conclusion} concludes the paper.

\section{Preliminaries and Related Work\label{sec:related}}

BE-RAN incorporates key RAN features and blockchain concepts to implement core network functions at the RAN level, laying a foundation for future 6G networks. While current 5G architectures form the basis of our research, the principles and innovations proposed in BE-RAN are designed with forward compatibility in mind, anticipating the evolving needs of 6G and beyond \cite{Xu20206G}.

Current 5G RAN typically comprises Centralized Units (CU), Distributed Units (DU), and Radio Units (RU) operating at different OSI layers \cite{O-RANAlliance2020}, with the RAN Intelligent Controller (RIC) operating at upper OSI layers. This disaggregated architecture is expected to evolve further in 6G, potentially introducing new functional splits and increased intelligence at the edge. BE-RAN's flexible approach to identity management and authentication is designed to accommodate these future developments, aligning with the vision of decentralized radio access networks for 5.5G and beyond \cite{xu2023decentran}.
Blockchain uses consensus mechanisms and linked blocks to maintain data integrity \cite{nakamoto2008}. BE-RAN introduces a global blockchain address (BC ADD) as a universal identifier for UEs and RAN elements, enabling user-centric identity management and mutual authentication. This approach builds upon previous work in blockchain-enabled resource management for 6G communications \cite{Xu20206G}, extending the concept to identity and authentication processes. Unlike B-RAN \cite{Ling2020} and BC-RAN \cite{Tong2020}, BE-RAN uses blockchain solely for identity exchange, applying zero-trust principles throughout.

BE-RAN's mutual authentication mechanism uses BC ADDs, improving security without relying on vulnerable centralized Certificate Authorities \cite{Zwarico2020}. This approach reduces identity-based attack risks and eliminates single points of failure in the RAN.

In terms of service logging and billing, BE-RAN leverages blockchain for RAN-level logging services, potentially supporting cross-operator transactions. This approach offers a more flexible, decentralized, and user-centric billing system compared to traditional core network processing. The concept of blockchain-enabled decentralized name services and P2P communication protocols \cite{zhou2023be} further supports this vision of a decentralized and user-centric network infrastructure.

\section{Blockchain-enabled RAN\label{sec:arch}}

BE-RAN adapts conventional RAN with a decentralized focus, enabled by novel identity management and mutual authentication. It employs a top-down approach for function deconstruction and a bottom-up approach for deployment, ensuring lower layers function independently from upper-level RAN and CN elements.

BE-RAN's key features include identity management and mutual authentication at the UE level, improving locality and latency. It enables ad hoc network switching for UE-to-UE authentication within local DU groups. RAN elements utilize BC ADD association, complementing conventional ID and PKI-based credentials. For out-of-reach DUs, BE-RAN implements upper-layer traffic routing with specific headers. The architecture also integrates with MEC at the network layer.

When operating predominantly in the BE-RAN regime, RAN effectively becomes an Edge Network. This transformation allows RIC to implement sophisticated services and policies such as zero-rated service, QoS, and QoE for connected UEs and RAN elements.

% \label{sec:case}
\begin{figure}[ht]
  \centering
  \includegraphics[scale = 0.31]{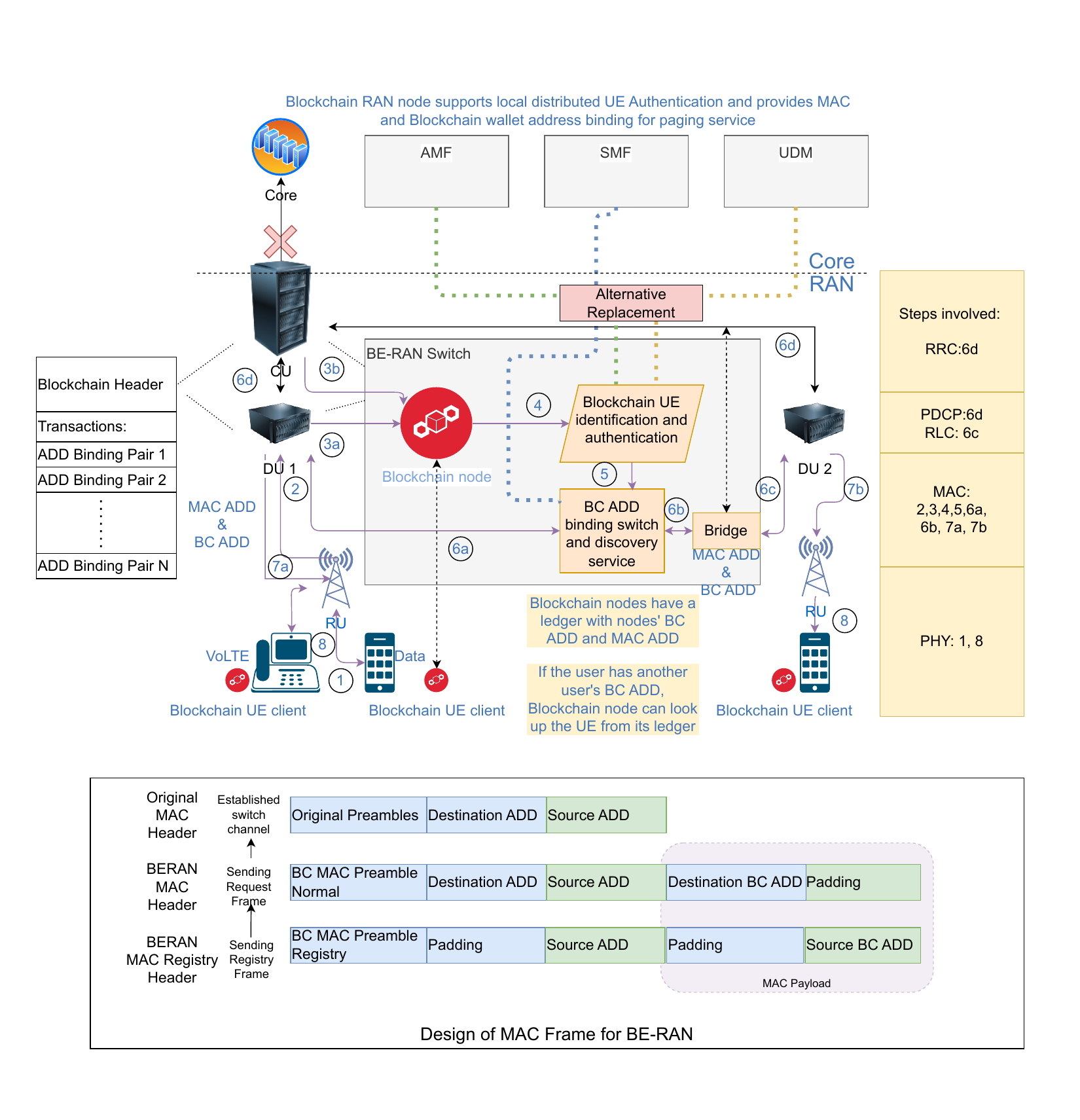}
  \caption{Framework and Frame structure of BE-RAN at Layer 2/MAC layer \label{fig:case}}
\end{figure}

\subsection{Framework}
The BE-RAN architecture consists of top-level elements such as RICs, and lower-level RAN elements including DUs and CUs. This structure aligns with the basic RAN functional split option 7.2 defined by 3GPP \cite{Sirotkin2020}, where RUs, DUs, and CUs are organized hierarchically. BE-RAN is designed to support local groups of users who share the same RU, DU pool, and CU, enabling efficient resource utilization and management. A detailed case study illustrating this concept is presented in later sections, as shown in Fig. \ref{fig:case}.
The blockchain usage spans all BE-RAN elements except RU due to a lack of coded information at the RU PHY interface. The services of BE-RAN are roughly classed into four levels starting with BE-RAN User/UE registry to local and regional switching and routing, powered by BeMutual protocols, illustrated in Fig. \ref{fig:mutual}.

The distributed BE-RAN architecture also takes advantage of the virtualized hosting of logical RAN elements by making the blockchain (BC) node a communal node for all RAN logical units. It also benefits from RICs for policing and QoS potential by influencing RIC and administration of control panel interface. It is worth pointing out that the communication link can be set up between UE-UE, UE-RAN element, RAN element-RAN element, etc., as long as the interfaces are given BC-ADDs. The framework suggested in this paper does not restrict the network topology or the roles of users in more general considerations, but the paper is scoped to cover RAN as much as possible.

\subsection{Entities and Functions}

BE-RAN consists of three primary components: UE, RAN elements, and BC nodes. The BC nodes in RAN elements are more powerful than those in UEs, but all use the same distributed ledger with blockchain protocols.

Key entities include:

\begin{itemize}
  \item UE: hosting a lightweight BC node that handles BE-RAN MAC frames and packets. Uses BC ADD as its identity.
  
  \item Blockchain node: a logical node hosted in UE or RAN elements, maintains a shared ledger and synchronizes with other nodes.
  
  \item BC ADD: a user-generated encrypted address serves as anonymous identity.
  
  \item BE-switch: a modified switch reading blockchain-enabled MAC Frames, matching BC-ADD and MAC ADD.
  
  \item BE-router: a modified router forwarding traffic based on BC-ADD and IP address pairing.
\end{itemize}

\subsection{Stakeholders and Services}

\begin{itemize}
\item Users: UEs or RAN elements with blockchain clients, identified by BC ADD.

\item RAN elements: Act as miners, communicating via X2/E2 interfaces.

\item Mutual authentication: Achieved through BC-ADD holders' private key signatures.

\item Privacy preservation: Implemented via anonymous BC ADD associated with physical addresses.

\item Routing and switching: Utilizes UE attachment encapsulation and altered frame structures.

\item Billing/Logging: a logging service enabled through blockchain, with interactions logs via encrypted address with UTXO or accounts.

\item User legitimacy: Determined by BC ADD and its balance.
\end{itemize}

\section{BE-RAN security and privacy-preserving communication\label{sec:func}}

\subsection{Blockchain-enabled Mutual Authentication (BeMutual)}

BeMutual leverages blockchain to enable user-centric, distributed mutual authentication \cite{Wang2010}. It binds global blockchain addresses (BC ADD) with local physical addresses (ADD), complementing traditional authentication methods while eliminating the need for third-party trust.

\begin{figure}[h]
  \includegraphics[width = 0.48\textwidth]{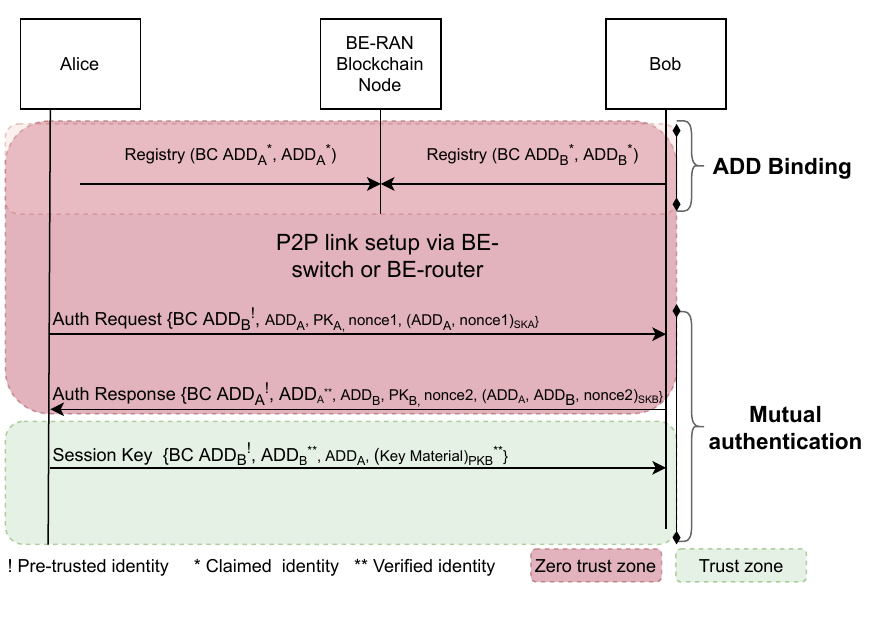}
  \caption{Mutual authentication of BE-RAN \label{fig:mutual}}
\end{figure}

The core of BeMutual is the one-way relationship between public key (PK) and BC ADD. BC ADD is generated by hashing PK, allowing easy verification of claimed bindings while preventing forgery. This process is illustrated in Fig. \ref{fig:mutual}'s registry message.

Due to the zero-trust model, a two-step authentication process is implemented. First, UEs/RAN elements register their BC ADD-ADD binding with the BE-RAN blockchain node. Second, mutual authentication is performed between two entities to verify each other's binding and generate a session key for secure end-to-end communication. Fig. \ref{fig:mutual} details this process between Alice and Bob.

This approach ensures strong identity security among peers without relying on centralized authorities, making it applicable not only for RAN authentication but also for broader scenarios requiring distributed identity management.

\begin{enumerate}
  \item Alice $\rightarrow$ Bob: Authentication Request 
  
  Alice sends the pre-trusted $BC ADD_B$, its address $ADD_A$, its public key $PK_A$, and a random number $nonce1$ with a signature of $(ADD_A, nonce1)$ by using its private key $SK_A$ to Bob. The $nouce1$ is introduced to prevent multiple types of attacks on integrity and prevent man-in-the-middle threats.
  
  \item Bob $\rightarrow$ Alice: Authentication Response

  Upon reception of the Authentication Request from Alice in 1), Bob first checks if the claimed $ PK_A$ can derive $ BC ADD_A$. If it matches pre-trusted $BC ADD_A$, the claimed $PK_A$ belongs to $BC ADD_A$. In the next step, Bob uses $PK_A$ to verify the signature signed by Alice's private key, noted by $SK_A$, to check if the decrypted $nonce1$ from the signature is equal to the claimed $nonce1$, to prevent replay attacks, so is the $ADD_A$. Once Alice's authentication is done, Bob can believe this message is from $BC ADD_A$ because of matched records in both plain text and encrypted text signed by Alice's private key. Then, Bob constructs an Authentication Response to Alice.

  \item Alice $\rightarrow$ Bob: Session Key Confirmation

  After receiving the Authentication Response from Bob, detailed in 2), Alice checks if this message is from Bob by validating $PK_B$ and $BC ADD_{B}$, with one more verification of $ADD_B$ by comparing the plain text and decrypted $ADD_B$ from $SK_B$ encrypted message. Then, it chooses a session key or generates a session key by using the $nonce1$ and $nonce2$ for further key negotiation and sends the session key confirmation message to Bob. 
  
  \item Once the session key has been exchanged, a secured communication channel is set up with the proposed mutual authentication protocol. 
\end{enumerate}

By completing mutual authentication without third-party involvement, users' privacy is kept to their own with easy identity management, as the BC ADD can be kept the same as a mobile phone number or e-mail address in the era of decentralization.

BeMutual is a great help to solve the identity crisis in cybersecurity and communication security. RAN elements with BE-RAN can easily verify other RAN elements within or among other RAN networks, as the RAN elements are now globally authenticated. A great use case of it would be stopping fake base stations. Once the UE has a list of trusted base stations from a trusted blockchain ledger maintained by communities and MNOs (the temper-proof and consensus will ensure the correctness of the record), it will never connect to unauthorized base stations.

\subsection{Independent RAN operation and alternative logging and billing on RAN}
As indicated in the proposed framework, while most traffic happens internally, the RAN is solely operating on its own with most service capabilities, such as voice/data service for the general public. It becomes a new operation model for MNOs and private or even virtual MNOs. 

Most importantly, BE-RAN enables D2D communication to be authenticated more locally than ever, as 3GPP defined D2D communication requires CN involvement due to the security concerns on identity authentication. 
In addition, billing and logging on RAN usage ideally resides on the blockchain function, where the operators may choose to make BE-RAN with tokens that are universal to all stakeholders, with strong initiatives opening up the sharing of RAN to other MNOs and private sectors. BE-RAN opens up new business models that will thrive both MNOs and their customers.

\subsection{Privacy-preserving P2P communication by RAN}

Traditional P2P communication faces challenges in achieving true privacy preservation due to the reliance on centralized infrastructure. Even in decentralized networks like Bitcoin \cite{nakamoto2008}, central routing services or third-party discovery mechanisms expose user identities and activities. BE-RAN addresses these privacy challenges through several innovative approaches. It implements user-centric identity management, where users interact using only their BC ADD, eliminating the need for centralized identity databases. The system employs decentralized authentication, with mutual authentication and integrated key negotiation enabling E2E encryption without third-party involvement. Furthermore, BE-RAN allows selective identity disclosure, where physical addresses are only used when claimed by the user, reducing the risk of unauthorized tracking.

While enhancing privacy, BE-RAN also considers regulatory compliance. Mobile Network Operators (MNOs) can implement measures such as BC ADD or IMEI whitelisting and mandatory Know-Your-Customer (KYC) registration to prevent misuse while maintaining user privacy \cite{xu2023decontroller}. This approach balances user privacy with necessary regulatory oversight, providing a more secure and private P2P communication framework within the RAN infrastructure.

\section{Results}\label{sec:results}

\begin{table*}[htpb!]
  \centering
  \caption{Comparison table of selected common mutual authentication protocols}
	\label{tab:com}
  \begin{adjustbox}{width=0.95\textwidth}
  \begin{tabular}{|c|c|c|c|c|c|c|c|c|c|c|}
    \hline
     \multicolumn{10}{|c|}{Communication Overhead} & Computation Overhead  \\
    \hline
     & signal 1 (bits) & signal 2 (bits) & signal 3 (bits)& signal 4 (bits) & \multicolumn{5}{c|}{signal 5-9 (bits)}  &  \\
    \hline
    BE-RAN  & \begin{tabular}[c]{@{}c@{}}784\footnotemark[1]+2ADD+ \\ PK \end{tabular}  & \begin{tabular}[c]{@{}c@{}}784\footnotemark[1]+4ADD+ \\ PK \end{tabular} &0  & 0 &  0& 0 & 0 & 0 & 0  & $2T_{sign}+2T_{verify}+2T_{hash}$ \\
    \hline
    IKEv2 & \begin{tabular}[c]{@{}c@{}c@{}}(EC)DHPara+\\nonce  \end{tabular}& \begin{tabular}[c]{@{}c@{}c@{}}(EC)DHPara+\\nonce  \end{tabular} & \begin{tabular}[c]{@{}c@{}c@{}}2ADD+\\2CERT+\\nonce+prf() \end{tabular} &\begin{tabular}[c]{@{}c@{}c@{}}2ADD+\\2CERT+\\ nonce+prf() \end{tabular}  & 0 & 0 & 0 & 0 & 0  & \begin{tabular}[c]{@{}c@{}c@{}}$T_{(EC)DH}+4T_{sym}+$\\$4T_{hmac}+2T_{sign}+$\\$2T_{verify}$ \end{tabular} \\
    \hline
    TLS 1.3 &\begin{tabular}[c]{@{}c@{}c@{}}(EC)DHPara+\\nonce \end{tabular}&\begin{tabular}[c]{@{}c@{}c@{}}(EC)DHPara+\\nonce \end{tabular}&ignored\footnotemark[2]&\begin{tabular}[c]{@{}c@{}}CERT \\ or PK \end{tabular} &\begin{tabular}[c]{@{}c@{}}hash\\(sign.)\end{tabular}&hmac&\begin{tabular}[c]{@{}c@{}}CERT \\ or PK \end{tabular}&\begin{tabular}[c]{@{}c@{}}hash\\(sign.)\end{tabular}&hmac&\begin{tabular}[c]{@{}c@{}}$T_{(EC)DH}+14T_{sym}+2T_{sign}+$\\ $2T_{hash}+2T_{verify}+2T_{hmac}$ \end{tabular}\\
    \hline

  \end{tabular}
\end{adjustbox}

  \end{table*}

  \begin{table}[h]
    \caption{Table of parameters}
    \label{tab:para}
    \begin{adjustbox}{width=0.45\textwidth}
    \begin{tabular}{|c|c|c|}
      \hline
      Abbr. & Name &Length\footnotemark[3]\\
      \hline
      IKEv2 & Internet Key Exchange v2 & N/A\\
      \hline
      TLS & Transport Layer Security & N/A\\
      \hline
      (EC)DHPara & Elliptical Curve DH parameters & 256 bits\\
      \hline
      DHPara &DH parameters & 3072 bits\\
      \hline
      nonce & A nonce by prf()& 256 bits\\
      \hline
      ADD & IPv6 address & 128 bits\\
      \hline
      prf() & Output of Pseudo-random functions & =256 bits\\
      \hline
      CERT & X.509v3 Certificate(s)&  5592 bits\\
      \hline
      PK (RSA/DSA) & Public key with RSA or DSA & 3072 bits\\
      \hline
      SK (RSA/DSA) & Private key with RSA or DSA & 256 bits\\
      \hline
      PK (ECDSA) & Public key with ECDSA& 256 bits\\
      \hline
      SK (ECDSA) & Private key with ECDSA & 256 bits\\
      \hline
      hash(sign.) & A hashed signature using SHA256 &256 bits\\
      \hline
      hmac & A hmac value & 256 bits\\
      \hline
      $T_{pm}$&Time of a scale multiplication& 0.906 ms\\ 
      \hline
      $T_{exp}$&Time of a modular exponentiation & 0.925 ms\\
      \hline
      $T_{hash}$ &Time of a hash operation & 0.5 us\\
      \hline
      $T_{signR}$ & Time of a sign. operation for RSA& 1.506 ms\\
      \hline
      $T_{verifR}$ & Time of a verfiy operation for RSA & 0.03ms\\
      \hline
      $T_{signE}$ & Time of a sign. operation for ECDSA& 0.016 ms\\
      \hline
      $T_{verifE}$ & Time of a verify operation for ECDSA & 0.1 ms\\
      \hline
      $T_{sym}$ & Time of a symm. key operation & 3 us\\
      \hline
      $T_{hmac}$ &Time of a hmac operation & 1.4 us\\
      \hline
      $T_{DH}$ &Time of a DHPara operation &1.812 ms\\
      \hline
      $T_{ECDH}$ &Time of a ECDHPara operation &2.132 ms\\
      \hline

    \end{tabular}
  \end{adjustbox}
  \end{table}

As the BE-RAN is a framework with two basic function groups, it requires attention to performance on mutual authentications. The results of BE-RAN are presented regarding the communication and computation metrics. The comparable protocol stacks are commonly found in the Transportation Layer or Network Layer. We have selected a few iconic stacks for benchmarking purposes, they are Internet Key Exchange version 2 (IKEv2) \cite{Pagliusi2004} and Transport Layer Security (TLS) \cite{Holz2019} with multiple cryptography methods.

\subsection{Performance analysis of mutual authentication protocols}

This analysis compares BeMutual with common cellular network practices: certificate-based IKEv2 and TLS 1.3 (both certificate-based and public-key-based) \cite{Zwarico2020}. The comparison focuses on signaling, communication, and computation aspects, excluding optional parameters and payloads for simplicity (see Table \ref{tab:com}).

IKEv2 consists of two phases: Initial Exchange ($IKE-INIT$) and Authentication Exchange ($IKE-AUTH$). $IKE-INIT$ involves two messages containing nonce and $(EC)DH$ parameters to calculate $KEYSEED$. $IKE-AUTH$ messages include sender's ID, certificate, and an $AUTH$ value derived from these components plus a nonce and Pseudo-Random Function output.

TLS 1.3 handshake involves multiple steps, including Client\_Hello, Server\_Hello, certificate exchanges, and verifications. Client\_Hello and Server\_Hello contain nonce and $(EC)DHE$ parameters to generate a handshake secret. Subsequent messages are encrypted using keys derived from this secret. Certificate\_Verify provides a signature over the entire handshake, while Finished messages contain a MAC value for integrity.

\paragraph{Overhead Analysis}
BeMutual requires only 2 signals, compared to 4 for IKEv2 and 9 for TLS 1.3. Environment overhead, particularly for BE-RAN, depends on the underlying blockchain platform and consensus mechanism. For static BE-RAN deployment, user BC ADD synchronization is assumed, minimizing additional costs.

\paragraph{Communication Overhead}
Based on TLS 1.3, IKEv2, and NIST standards, we assume 256-bit nonces and hash outputs, 3072-bit public keys for finite-field cryptography, 256-bit private keys, and 256-bit keys for ECC. BC ADD length is assumed to be 272 bits. X.509 v3 certificates are 699 bytes, and IPv6 addresses (128 bits) replace IKEv2 IDs.

\paragraph{Computation Overhead}
We define time variables for various cryptographic operations (e.g., $T_{sym}$, $T_{sign}$, $T_{verify}$, etc.). Benchmark tests use 1024-byte plaintexts and 5592-bit certificates.

\begin{figure}[h]
  \centering
  \subfloat[Communication overheads for Finite field and ECC crypto protocols]{
    \includegraphics[width=0.23\textwidth]{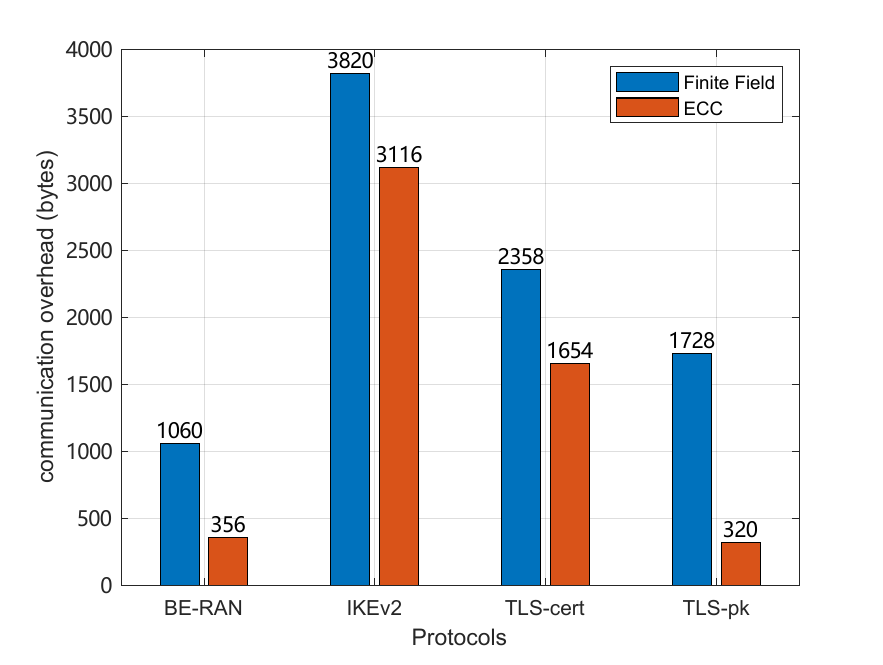}
    \label{fig:commcost}
  }
  \hfill
  \subfloat[Computational overhead for Finite field and ECC crypto protocols]{
    \includegraphics[width=0.23\textwidth]{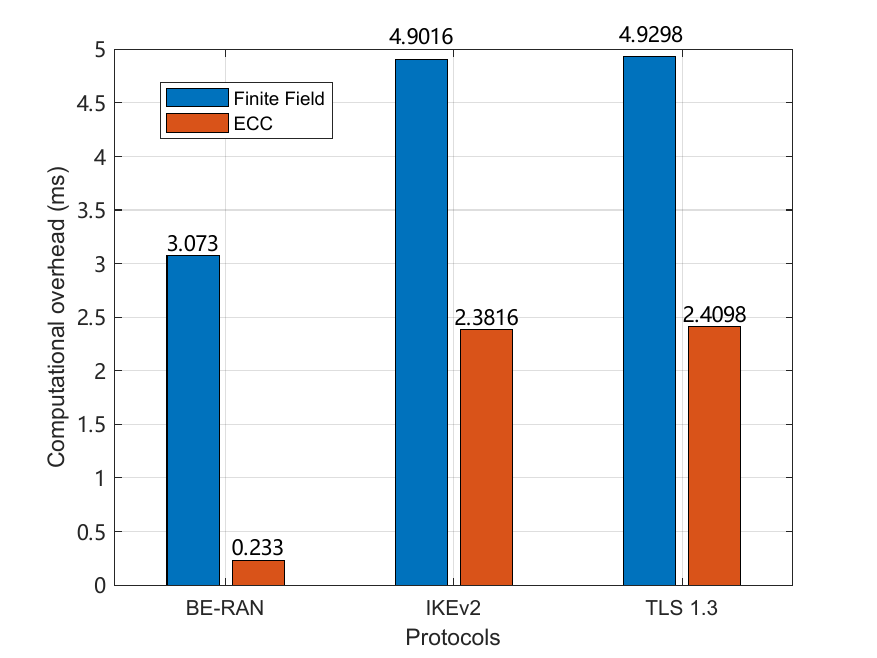}
    \label{fig:computecost}
  }
  \caption{Performance comparison of Finite field and ECC crypto protocols}
\end{figure}

\subsection{Performance Comparisons}
We compare BeMutual with IKEv2 and TLS 1.3, using both RSA and ECDSA \cite{Pagliusi2004}. Fig. \ref{fig:commcost} shows BE-RAN outperforms in Finite Field-based algorithms and matches TLS in ECC-based algorithms (356 vs. 320 bytes).

The computational overhead is measured by the execution time of protocols, conducted on Ubuntu 20.04.1 4GB RAM, with 4 Cores at 4.2 GHz virtualized on a PC with Windows 10 installed. With the same physical environment, the computational overhead is measurable by comparing the execution time on the same platform, listed in Table. \ref{tab:para}. Selected authentication protocols are IKEv2 with RSA/ECDSA, TLS 1.3 with RSA/ECDSA and BE-RAN with RSA/ECDSA.

The benchmarked results of all computational components are listed in Table \ref{tab:para}, and the comparison of selected protocols are shown in Fig. \ref{fig:computecost}. The result of the computational comparison shows BE-RAN has significant improvements on Finite-Field-based algorithms, e.g., RSA and DSA, and outperformed other protocols by a considerable margin with the elliptical curve method. The ECC-based results provide a good foundation for BE-RAN to be used in IoT and other thin-clients due to the lighter authentication cost.
  
\section{Challenges and Future work\label{sec:challenges}}

BE-RAN introduces challenges and opportunities for future research. Expanding BeMutual offers potential for a widely applicable zero-trust model for distributed identity management, though scalability and performance challenges exist when implementing blockchain across networks.

Integrating blockchain into lower OSI layers poses difficulties, particularly for DUs with limited computing capabilities, necessitating modifications to RAN control plane interfaces. Developing standardized protocols for BE-RAN switching and expanding routing capabilities are crucial next steps, including ensuring packet compatibility and integration with existing infrastructure.

BE-RAN presents opportunities to transform base stations into wireless boosters for local controllers, benefiting applications like autonomous vehicles and IoT devices. Future research should address user-centric policies, network health supervision, and traffic management. Blockchain-based mechanisms could be leveraged for access control, logging and billing, potentially integrating with RAN Intelligent Controllers.

\section{Conclusions\label{sec:conclusion}}

We proposed a blockchain-enabled radio access network and its full potential to power a decentralized privacy-preserving P2P communication system with secured identity management via mutual authentication. Detailed authentication protocols and implementation examples are given, and the proposed BE-RAN system has better performance than the common practice of CA or public key PKI in terms of communication and computation overheads.

\normalem
\bibliographystyle{IEEEtran}                                
\bibliography{BERAN-arch}

\end{document}